\documentclass[aps,pra,]{revtex4-1}

\usepackage{amssymb}
\usepackage{amsthm}
\usepackage{amsmath}
\usepackage{graphicx,color}
\usepackage{acronym}
\usepackage{amsfonts}
\usepackage{amscd}
\usepackage{dcolumn}   
\usepackage{bm}        
\usepackage{braket}
\usepackage{natbib}

\usepackage[colorlinks=true]{hyperref}

\newcommand{\HH}{{\cal H}}
\newcommand{\A}{{\cal A}}

\begin{document}

\title{\bf How entangled can a multi-party system possibly be?}

\author{Liqun Qi}
\email{maqilq@polyu.edu.hk}
\affiliation{Department of Applied Mathematics, The Hong Kong Polytechnic University, Hong Kong}

\author{Guofeng Zhang}
\email{magzhang@polyu.edu.hk}
\affiliation{Department of Applied Mathematics, The Hong Kong Polytechnic University, Hong Kong}

\author{Guyan Ni}
\email{guyan-ni@163.com}
\affiliation{College of Science, National University of Defense Technology, Changsha 410073, Hunan, China}

\date{\today}

\begin{abstract}

The geometric measure of entanglement of a pure quantum state is defined to be its distance to the space of product (seperable) states. Given an $n$-partite system  composed of subsystems of dimensions $d_1,\ldots, d_n$, an upper bound for maximally allowable entanglement is derived in terms of geometric measure of entanglement. This upper bound is characterized exclusively by the dimensions $d_1,\ldots, d_n$ of composite subsystems. Numerous examples demonstrate that the upper bound appears to be reasonably tight.

%

\end{abstract}

\pacs{03.67.Mn, 03.65.Ud, 03.67.-a,}

\maketitle

\section{Introduction}

The physical realization of quantum computing calls for a hierarchical quantum network. The bottom level is the one- and two-qubit regime, where a photon interacts with matter (e.g., a trapped ion). In this regime, precise control must be exerted. Going one level up we enter the regime of quantum logic gates where typically ten or more qubits operate. One level further up is the fault-tolerant quantum error correction (QEC) architecture regime where hundreds of qubits reside. The final level is the algorithms regime.  Being an essential resource for quantum computing, entanglement propagates over the  dynamic quantum network to fulfil desired quantum computing  tasks. A fundamental quantum naturally arises: how much entanglement can a quantum network encode?

If a quantum network is composed of qubits, that is, each particle lives in a two-dimensional Hilbert space, we end up with a multipartite qubit system. When restricted to the pure state case,  the 2-qubit entanglement is well-understood. For the 3-qubit case,  it is well-known that the GHZ state \cite{GHZ89} is the most entangled  state in terms of entanglement entropy and its degree of entanglement can also be easily computed by means of many other measures of entanglement. On the other hand, it has been reported \cite{TWP09}  that the 3-qubit W-state  \cite{DVC00} is more entangled than the 3-qubit GHZ state in terms of geometric measure of entanglement \cite{WG}.  In fact, it is generally agreed that the characterization and quantification of entanglement of $n$-qubit systems for $n> 3$ is a difficult task.  In addition to qubits, for a typical quantum network,  there may also exist other finite-level units, see several recent experimental set-up in e.g.,  \cite{LWL+08, MEH+16, KMF+16, KRR17}. For such a hybrid quantum network, namely a heterogeneous multipartite system, it is unclear how much entanglement can be allowed, not to mention how to quantify it efficiently.

A fundamental problem in quantum physics and also an important problem in quantum information science is to detect whether a given state is entangled, and if so, how entangled it is.  Several measures of quantum entanglement have already been proposed in the literature, e.g., Schmidt rank  \cite[Section 2.5]{NC},  von Neumann entropy \cite[Section 11.3]{NC}, entanglement of formation \cite{BDS+96, *W98}, quantum concurrence  \cite{HW97, CKW00}, the Peres-Horodecki criterion \cite{HHH96, Peres96}, Schmidt measure (also called Hartley entropy) \cite{EB01} based on Candecomp/Parafac (CP) decomposition of tensors \cite{BG09, QL17}, relative entropy \cite{VP98},   negativity \cite{ZHS+98}, the geometric measure of entanglement, \cite{Shimony95, BL01, WG,  HJ09, CXZ, HS, NQB14, HQSZ14, HQZ16, BBG}. More can be found in the survey papers \cite{PV07, AFO+08, HHH+09}. For the bipartite pure state case, a state is maximally entangled  in terms of one measure is often also maximally entangled in  terms of another measure.  In this sense, different measures give consistent prediction. This is not true for multipartite cases. For a multipartite system, it is typical that two different measures attain their maxima at different quantum states \cite{DVC00, EWZ17}.

In this paper we are interested in the following problem: Given an $n$-partite system which can be either homogeneous or heterogeneous, how entangled can its states be? We will use the geometric measure to quantify the degree of entanglement.  We show that an upper bound can be derived for entanglement content allowed. Moreover, the upper bound is given exclusively in terms of dimensions of the composite subsystems. Not surprisingly, the upper bound can always be reached in the case of bipartite systems.  Interestingly, various examples demonstrate that upper bounds appear to be reasonably tight for many multipartite systems. 


\section{Geometric measure of entanglement (GME)} \label{sec:GME}

For a quantum $n$-partite system, a pure state $|\Psi\rangle$ is an element in the tensor product Hilbert space ${\cal H}={\cal H}_1\otimes \cdots \otimes {\cal H}_n \equiv \otimes_{k=1}^n {\cal H}_k$. For each  $k=1,\ldots, n$, denote the dimension of the composite subsystem ${\cal H}_k$ by $d_k$ and the orthonormal basis by  $\{ | e_{i_k}^{(k)} \rangle\}$. (For ease of presentation and without loss of generality, it is assumed in this paper that $d_1\leq d_2\leq
\cdots \leq d_n$.)  Then, a pure state $|\Psi\rangle  \in {\cal H}$ is of the from
\begin{equation}\label{eq:Psi}
| \Psi \rangle = \sum \bar{a}_{i_1\cdots i_n} |
e_{i_1}^{(1)} \rangle \otimes\cdots \otimes | e_{i_n}^{(n)} \rangle,
\end{equation}
 where $\bar{a}_{i_1\cdots i_n} \in \mathbb{C}$ (the ``bar'' stands for the  complex conjugation). The normalization condition of $\ket{\Psi}$ is $ \sum_{i_1, \cdots, i_n} |a_{i_1\cdots i_n}|^2=1$. A state $|\Phi\rangle \in {\cal H}$ is said to be {\it separable} if it is a product state
 \begin{equation} \label{july12_pure}
 |\Phi\rangle = |\phi^{(1)}\rangle \otimes\cdots \otimes |\phi^{(n)}\rangle,
 \end{equation}
 where
 \begin{equation} \label{eq:phi_k}
 |\phi^{(k)} \rangle = \sum u_{i_k}^{(k)} |
e_{i_k}^{(k)} \rangle \in {\cal H}_k, ~~~ \forall k=1,\ldots, n.
\end{equation}
 If a state is not separable, then it is called an {\it entangled} state.

Next, let us briefly review the geometric measure of entanglement. Denote the set of all separable pure states in $\HH$ as $Separ(\HH)$. For a general $n$-partite state $|\Psi \rangle \in \HH$, the geometric
measure of its entanglement content can be defined as its distance to the space of separable states $Separ(\HH)$, \cite{WG,HQZ16}, i.e.,
\begin{equation} \label{near}
d \triangleq \min
\left\{ \left\| |\Psi \rangle - |\phi \rangle \right\| : |\phi
\rangle \in Separ(\HH) \right\}.
\end{equation}
 Since the minimization in (\ref{near})
is taken with a continuous function on a compact set  $Separ(\HH)$ in a finite
dimensional space $\HH$, the minimizer does exist and is denoted by  $|\phi_\Psi \rangle \in Separ(\HH)$. Clearly,  $|\phi_\Psi \rangle$ is the  separable
state which is closest to $|\Psi \rangle$.

For convenience, as in \cite{WG,HQZ16}, instead of computing (\ref{near}) directly,
we study
\begin{eqnarray}
d^2
&=& \left\| |\Psi \rangle - |\phi_\Psi \rangle \right\|^2
\nonumber
\\
&=&
 \min
\left\{ \left\| |\Psi \rangle - |\phi \rangle \right\|^2 : |\phi
\rangle \in Separ(\HH) \right\}.
\label{near1}
\end{eqnarray}
Note that
\begin{equation*} \label{more}
\left\| |\Psi \rangle - |\phi \rangle \right\|^2
 = 2 - \langle \Psi |\phi
\rangle - \langle \phi |\Psi \rangle.
\end{equation*}
 Thus the
minimization problem in (\ref{near1}) is equivalent to the following
maximization problem:
\begin{equation} \label{near2}
\max_{\langle \phi^{(k)} |\phi^{(k)} \rangle = 1, k=1, \cdots, n}
\left\{ \langle \Psi | \otimes_{k=1}^n |\phi^{(k)}
\rangle  + \otimes_{k=1}^n \langle \phi^{(k)} |
 \Psi \rangle \right\}.
\end{equation}

Introducing Lagrange multipliers $\lambda_k$, $k = 1, \cdots, n$, and applying complex differentiation \cite{Huy05} to get
\begin{equation*} \label{e1}
\langle \Psi | \otimes_{j=1, j \neq k}^n |\phi^{(j)} \rangle
= \lambda_k \langle \phi^{(k)} |,
\end{equation*}
and
\begin{equation*} \label{e2}
 \otimes_{j=1, j \neq k}^n \langle \phi^{(j)}  | \Psi
\rangle= \lambda_k | \phi^{(k)} \rangle.
\end{equation*}
Therefore,
\[
 \lambda_k = \langle \Psi | \phi \rangle = \langle
\phi | \Psi \rangle, ~~ k=1,\ldots, n
\]
 is a real number in the interval $[-1, 1]$. The maximal overlap is,  \cite{WG},
\begin{equation} \label{near3}
\langle \Psi | \phi_\Psi \rangle \triangleq \max \{ | \langle
\Psi | \phi \rangle | : \phi \in Separ(\HH) \}
\end{equation}
and the geometric measure of entanglement of $|\Psi\rangle$, defined in \eqref{near}, is hence
\begin{equation} \label{july12_17}
d=\sqrt{2-2\langle \Psi | \phi_\Psi \rangle}.
\end{equation}
Clearly, the smaller the maximal overlap $\langle \Psi | \phi_\Psi \rangle$ is, the bigger the distance $d$ between $|\Psi\rangle$ and the set of separable states.

Next, we represent the geometric measure of entanglement by means of tensor  (also called hypermatrix) \cite{BG09, QL17}.  For the pure state $|\Psi\rangle$ in (\ref{eq:Psi}),  we define an associated tensor  $\A_\Psi$ by $\A_\Psi=
\left(a_{i_1\cdots i_n}\right) \in \mathbb{C}^{d_1\times \cdots\times d_n}$. That is, we store all the probability amplitude of the state $|\Psi\rangle$ into a multi-array.  Similarly, we associate each $|\phi^{(k)}\rangle$ in \eqref{eq:phi_k}  with a column vector $u^{(k)} \in \mathbb{C}^{d_k}$, $k=1,\ldots,d_n$. Then we define a c-number
\begin{equation} \label{eq:tensor_vector}
\A_\Psi u^{(1)}\cdots u^{(n)} \triangleq \sum a_{i_1\cdots i_n}u^{(1)}_{i_1}\cdots u^{(n)}_{i_n}.
\end{equation}
The inner product between $|\Psi\rangle$ in \eqref{eq:Psi} and $|\Phi\rangle$ in  \eqref{july12_pure}   can be re-written as
\begin{equation}
\langle \Psi | \Phi \rangle = \A_\Psi u^{(1)}\cdots u^{(n)}.
\end{equation}
Denote the spectral radius of the tensor $\A$ by
\begin{equation} \label{spectral}
\sigma(\A_\Psi) \triangleq \max_{\| u^{(k)} \|^2 = 1,\
 k= 1, \cdots, n} \left\{ |  \A_\Psi u^{(1)}\cdots u^{(n)}| \right\}.
\end{equation}
Then the largest overlap in \eqref{near3} can be expressed as
\begin{equation} \label{bd}
\langle \Psi | \phi_\Psi \rangle =\sigma(\A_\Psi).
\end{equation}
As a result, the geometric measure of entanglement of the multi-partite state $|\Psi\rangle$ is
\begin{equation}\label{july12_d_2}
d=\sqrt{2-2\sigma(\A_\Psi}).
\end{equation}

In the literature of tensor optimization, several algorithms have been developed for computing the spectral radius of a given tensor $\A$. When $\A$ is symmetric, it can be proved that the spectral radius can be obtained when $u^{(1)}=\cdots =u^{(n)}$, \cite{HKW09}. In particular, if further $\A$ is real and with all nonnegative elements, then the spectral radius is given by its largest  Z-eigenvalue \cite{Qi05, HQZ16}.  In general, the spectral radius of a symmetric tensor $\A$ is its largest unitary symmetric eigenvalue (US-eigenvalue) \cite{NQB14}. An algorithm has been developed to find the largest US-eigenvalue of a given symmetric tensor \cite[Algorithm 4.1]{NB16}. When $\A$ is non-symmetric, its spectral radius is its largest unitary eigenvalue (U-eigenvalue) \cite{NQB14}. The algorithm proposed in \cite{NB16} can be modified to find the largest U-eigenvalue of a given non-symmetric tensor, see the algorithm in Appendix. All the examples in this paper are computed using these two algorithms. 


\section{The theoretical upper bound}

In this section, a theoretical upper bound is proposed for entanglement possibly allowed by a given $n$-partite system.

Define
\begin{equation}
\sigma \triangleq \min \left\{ \langle \Psi | \phi_\Psi
\rangle :  | \Psi \rangle \in \HH, \langle \Psi |
\Psi \rangle = 1 \right\},
\end{equation}
where $\ket{\phi_{\Psi}}$ is the the product state closest to $\ket{\Psi}$, as defined in the paragraph below \eqref{near}.  
Clearly, for any pure state $|\Psi \rangle \in \HH$,  we have
\begin{equation}
\|\Psi \rangle - |
\phi_\Psi \rangle \| \le \sqrt{2-2\sigma}.
\end{equation}
 So,  $\sqrt{2-2\sigma}$ is an upper bound of possible entanglement allowed in an $n$-partite system.  In what follows we give an estimate of this upper bound.

For a given $d_1\times d_2\times \cdots \times d_n$ tensor $\A$, by fixing the first $n-2$ indices $i_1, \cdots, i_{n-2}$, we end up with a  $d_{n-1}\times d_n$ matrix $A_{i_1\cdots i_{n-2}} \equiv ( a_{i_1\cdots i_{n-2}jk})$.   According to
(\ref{spectral}),
\begin{equation}
\sigma(A_{i_1\cdots i_{n-2}}) \le \sigma(\A).
\end{equation}
Let $\| \A \|$ and $\| A_{i_1\cdots i_{n-2}} \|$ be the Frobenius
norm of $\A$ and $A_{i_1\cdots i_{n-2}}$ respectively, i.e.,
$$\| \A \| = \sqrt{\sum_{i_1, \cdots, i_n} | a_{i_1\cdots i_n} |^2},$$
and
$$\| A_{i_1\cdots i_{n-2}} \| = \sqrt{\sum_{i_{n-1},i_n} | a_{i_1\cdots i_n}
|^2}.$$  By singular value decomposition (SVD), for the matrix $A_{i_1\cdots i_{n-2}}$ defined above,  we have
\begin{equation}
\|A_{i_1\cdots i_{n-2}}\|^2 \le d_{n-1}\sigma(A_{i_1\cdots i_{n-2}}).
\end{equation}
Since $\langle \Psi | \Psi \rangle = 1$, we have $\| \A \| = 1$.
Putting all of these together, we get
\begin{eqnarray*}
 1 &=& \| \A \|^2 = \sum_{i_1,\cdots, i_{n-2}} \| A_{i_1\cdots i_{n-2}}
\|^2
\\
&\le& \sum_{i_1,\cdots, i_{n-2}} d_{n-1}\sigma(A_{i_1\cdots
i_{n-2}})^2
\\
&\le& \sum_{i_1,\cdots, i_{n-2}} d_{n-1}\sigma(\A)^2
\\
&=&
 d_1\cdots
d_{n-1} \sigma(\A)^2.
\end{eqnarray*}
That is,
\begin{equation} \label{eq:sigma_A}
\sigma(\A) \geq  1/\sqrt{d_1\cdots d_{n-1}}.
\end{equation}
Because this is true for all $\A$ with $\| \A \| = 1$, by the definition of $\sigma$, we have
$\sigma \ge 1/\sqrt{d_1\cdots d_{n-1}}$.
Therefore, the upper bound is
\begin{equation} \label{july12_d_3}
d\leq \sqrt{2-2/{\sqrt{d_1\cdots d_{n-1}}}}.
\end{equation}

In particular, for an $n$-qubit system, namely $d_k=2$ for all $k=1,\ldots, n$, (\ref{july12_d_3}) reduces to
\begin{equation}
d \leq  \sqrt{2-2/\sqrt{2^{n-1}}}.
\end{equation}

\section{Examples}
In the previous section, a theoretical upper bound has been proposed for entanglement possibly allowed in  any given $n$-partite system. A natural question is: Given a multi-partite system, can the upper bound be reached? If yes, how to find the state that gives the maximal entanglement? If the upper bound cannot be reached, how tight is it?

When $n=2$, namely the bipartite case, the tensor $\A$ reduces to a $d_1\times d_2$ matrix $A=(a_{ij})$. By \eqref{eq:sigma_A}, $\sigma(A) \geq 1/\sqrt{d_1}$. Then by \eqref{july12_d_3},  $d \leq \sqrt{2-2/\sqrt{d_1}}$, i.e., the geometric measure of entanglement of bipartite pure states is no large than $\sqrt{2-2/\sqrt{d_1}}$.  It is well-known that this upper bound can always be reached. In fact, let $a_{jj} = 1/\sqrt{d_1}$ and $a_{ij} = 0$ if $i \not = j$. Clearly $\sigma(A) = 1 /\sqrt{d_1}$, which is attended  by the pure state
 \[
 | \phi_\Psi \rangle=\frac{1}{\sqrt{d_1}}\sum_{i=1}^{d_1} |e^{(1)}_i\rangle \otimes |e^{(2)}_i\rangle.
 \]
Readers may have recognized that the above procedure is essentially the Schmidt decomposition \cite{NC}, \cite{CHS00}.

Next, we look at several examples for various multi-partite cases, which indicate that the proposed theoretical upper bounds are often reasonably tight.

{\it Example 1: $3$-qubit system.} Given an $n$-qubit Greenberger-Horne-Zeilinger (GHZ) state, \cite{GHZ89}
\[
|n{\rm GHZ}\rangle = \frac{|0\rangle^{\otimes n} + |1\rangle^{\otimes n} }{\sqrt{2}},
\]
it is well-known that its geometric measure of entanglement is $\sqrt{2-2/\sqrt{2}}$, see e.g., \cite{WG, HQZ16}.
Clearly,
\begin{equation} \label{july13_2GHZ}
\sqrt{2-2/\sqrt{2}} \leq  \sqrt{2-2/\sqrt{2^{n-1}}}
\end{equation}
for all $n\geq 2$. Notice that the inequality \eqref{july13_2GHZ} is tight for $n=2$, the 2-qubit  case.  When $n=3$, namely the 3-qubit case, the geometric measure of entanglement of the $|3{\rm GHZ}\rangle $ state is strictly smaller than the upper bound which is 1 in this case. This means that the upper bound is not tight for this particular state. However, although $|3{\rm GHZ}\rangle $ is maximally entangled in terms of   3-tangle \cite{CKW00},   it is not the maximally entangled 3-qubit state in terms of the 2-tangle or the persistency of entanglement \cite{DVC00, BR01}.  A similar statement can be made for the geometric measure of entanglement.  In fact, as shown in \cite{WG, HQZ16}, the geometric measure of entanglement of the $|3{\rm GHZ}\rangle $ is $\sqrt{2-2/\sqrt{2}}\approx 0.7654$, while the geometric measure of entanglement of the W state \cite{DVC00}
\begin{equation}\label{july14_W}
|W\rangle = (|100\rangle + |010\rangle + |001\rangle)/\sqrt{3}
\end{equation}
is $\sqrt{2-2*2/3}\approx 0.8165$ with the closest symmetric product state
\[
\ket{\phi_\Psi} =  \ket{\phi}\otimes \ket{\phi}\otimes \ket{\phi},
\]
where \[\ket{\phi}=(-0.7885 + 0.2119 i)\ket{0} +(0.4996 + 0.2894 i)\ket{1}.\]

In fact, it is shown in \cite{TWP09} that the W state is the most entangled 3-qubit pure state in terms of the geometric measure of entanglement.  

{\it Example 2: 4-qubit system.} In this case, the theoretical upper bound is $1.1371$.  It is found that the GME of the state $\ket{\Psi} =\frac{1}{\sqrt{6}}(\ket{0011}+\ket{1100}+e^{2i\pi/3}(\ket{0101}+\ket{1010})+e^{4i\pi/3}(\ket{0110}+\ket{1001}))$ is $1.0282$ with the closest product state
 \[
\ket{\phi_\Psi} = \ket{\phi_1}\otimes\ket{\phi_2}\otimes\ket{\phi_3}\otimes\ket{\phi_4},\]
where
\begin{eqnarray*}
  \ket{\phi_1} &=& (-0.3674 + 0.3830 i)\ket{0}-( 0.6813 - 0.5042 i)\ket{1}, \\
  \ket{\phi_2} &=& (-0.9955 - 0.0569 i)\ket{0}-( 0.0418 + 0.0629 i)\ket{1}, \\
  \ket{\phi_3} &=& (0.5210 + 0.2240 i)\ket{0}-( 0.7665 - 0.3013 i)\ket{1}, \\
  \ket{\phi_4} &=& (0.6323 - 0.0502 i)\ket{0}-( 0.3042 + 0.7107 i)\ket{1}.
\end{eqnarray*}


{\it Example 3: 5-qubit system.}  In this case, the theoretical upper bound $1.2247$.   It is found that the GME of the following $5$-qubit absolutely maximally entangled state, \cite[(37)]{EWZ17},
\begin{eqnarray*}
\ket{\Psi} &=& \frac{1}{2\sqrt{2}}\left(\ket{00000}+\ket{00011}+\ket{01100}-\ket{01111}\right.\\
&&\left.+\ket{11010}+\ket{11001}+\ket{10110}-\ket{10101}\right)
\end{eqnarray*}
 is $1.1291$ with the closest product state
\[
\ket{\phi_\Psi} =\ket{\phi_1}\otimes\ket{\phi_2}\otimes\ket{\phi_3}\otimes\ket{\phi_4}\otimes\ket{\phi_5},
\]
where
\begin{eqnarray*}
  \ket{\phi_1} &=& (-0.7060 + 0.5388i)\ket{0}+(0.4556 + 0.0612i)\ket{1}, \\
  \ket{\phi_2} &=& (0.3766 + 0.8043i)\ket{0}+(  0.4322 + 0.1566i)\ket{1}, \\
  \ket{\phi_3} &=& (0.7843 + 0.4166i)\ket{0}+(0.1346 + 0.4395i)\ket{1},\\
  \ket{\phi_4} &=& (0.5652 + 0.6850i)\ket{0}-(0.4576 + 0.0439i)\ket{1}, \\
  \ket{\phi_5} &=& (0.2449 + 0.8536i)\ket{0}+(0.2228 - 0.4021i)\ket{1},
\end{eqnarray*}

{\it Example 4: 6-qubit system.}  In this case, the theoretical upper bound $1.2831$.   It is found that the GME of the $6$-qubit state in Equation (5) of Reference \cite{AC13} is $1.1927$.

{\it Remark 1.} The results in Examples 1-4 are summarized in Table \ref{table:qubits}.  It can be seen that  at least for $n\leq 5$,  there exists a state for an $(n+1)$-qubit system, whose GME can be very close to (slight higher or lower than) the theoretical upper bound  of entanglement possibly allowed by any $n$-qubit system.
\begin{table}[htp]
\caption{}
\begin{center}
\begin{tabular}{|c|c|c|}
\hline
$n$-qubit  & theoretical upper bound & best GME found\\
\hline
$2$-qubit  & 0.7654                    & 0.7654\\
\hline
$3$-qubit  & 1                   & 0.8165\\
\hline
$4$-qubit  & 1.1371                   & 1.0282\\
\hline
$5$-qubit  & 1.2247                    & 1.1291\\
\hline
$6$-qubit  & 1.2831                    & 1.1927\\
\hline
\end{tabular}
\end{center}
\label{table:qubits}
\end{table}%

{\it Example 5: 3-qutrit system.} In this case, the theoretical upper bound is $1.1547$. It can be verified that
 the GME of the 3-qutrit GHZ state
\begin{eqnarray*}
|\Psi\rangle =\frac{1}{\sqrt{3}}(|000\rangle  + |111\rangle +|222\rangle)
\end{eqnarray*}  
is $0.9194$  with the closest product state
$\ket{\phi_\Psi} = \ket{000}$. On the other hand, the GME of the 3-qutrit Dicke state $\ket{\Psi}= \frac{1}{\sqrt{6}}(\ket{012}+\ket{021}+\ket{102}+\ket{120}+\ket{201}+\ket{210})$ is $1.0282$  with the closest product state
\[
\ket{\phi_\Psi} =\ket{\phi}\otimes\ket{\phi}\otimes\ket{\phi},
\]
where $\ket{\phi}= (\ket{0}+\ket{1}+\ket{2})/\sqrt{3}$. Notice that this $3$-qutrit  Dicke state has the same entanglement content as the $4$-qubit state in Example 2. 
%
%
%

{\it Example 6: 4-qutrit system.}  In  this case, the theoretical upper bound $1.2709$. The GME of the following state, \cite[(B1)]{GZ14}, 
\begin{eqnarray*}
\ket{\Psi}  &=& \frac{1}{3} (\ket{0000}+\ket{0112}+\ket{0221}+\ket{1011}+\ket{1120}
\\
&&~~+\ket{1202}+\ket{2022}+\ket{2101}+\ket{2210})
\end{eqnarray*}
is $1.1547$ with  the closest product state
\[
\ket{\phi_\Psi} =\ket{\phi}\otimes\ket{\phi}\otimes\ket{\phi}\otimes\ket{\phi},
\]
where $  \ket{\phi} = (\ket{0}+\ket{1} +\ket{2})/\sqrt{3}$.


{\it Example 7: 4-ququart system.}    In  this case, the theoretical upper bound is  $1.3229$. The GME of the following 3-uniform state, \cite[(B4)]{GBZ16},
\begin{eqnarray*}
\ket{\Psi} &=& \frac{1}{4}\left(\ket{0000}+\ket{0123}+\ket{0231}+\ket{0312}\right.\\
&& +\ket{1111}+\ket{1032}+\ket{1320}+\ket{1203}\\
&&+\ket{2222}+\ket{2301}+\ket{2013}+\ket{2130}\\
&&+\left. \ket{3333}+\ket{3210}+\ket{3102}+\ket{3021}\right)
\end{eqnarray*}
is  $1.2247$,  with  the closest product state
\[
\ket{\phi_\Psi} =\ket{\phi}\otimes\ket{\phi}\otimes\ket{\phi}\otimes\ket{\phi},
\]
where
 $\ket{\phi}=-(i \ket{0}+i\ket{1}+\ket{2}-\ket{3})/2 $. Interestingly, the GME of this $4$-ququart system is the same as the theoretical upper bound of entanglement possibly allowed by all $5$-qubit systems.  It should be noted that the pure state $\ket{\Psi}$ is not symmetric.  Intuitively, the closest product state should also be non-symmetric. Interestingly, for this state, the largest U-eigenvalue can be obtained with the above product state $\ket{\phi_\Psi}$ which is symmetric.  The same is true for Example 6.

{\it Remark 2.} The results in Examples 2 and 6-7 are summarized inTable \ref{table:4-party}.   From Table \ref{table:4-party} we can see that the theoretical upper bound of entanglement allowed in all $4$-qubit systems can be overpassed by the GME of a particular $4$-qutrit state, and the theoretical upper bound of entanglement allowed in all $4$-qutrit systems can be approached by the GME of a particular $4$-ququart state.
\begin{table}[htp]
\caption{}
\begin{center}
\begin{tabular}{|c|c|c|}
\hline
$4$-party  & theoretical upper bound & best GME found\\
\hline
$4$-qubit  & 1.1371                     & 1.0282\\
\hline
$4$-qutrit  & 1.2709                   & 1.1547\\
\hline
$4$-ququart  &   1.3229                & 1.2247\\
\hline
\end{tabular}
\end{center}
\label{table:4-party}
\end{table}%
 
 {\it Example 8: $2\times 2\times 3$ system.}
 For this example,  the first two  particles  are qubits whereas third one is a qutrit.  The theoretical upper bound is 1. It can be verified that the GME of  the following 1-uniform state, \cite[(A2)]{GBZ16}, 
\begin{eqnarray*}
|\Psi\rangle =\frac{1}{\sqrt{6}}(\ket{000}+\ket{110}+\ket{011}+\ket{101}+\ket{002}-\ket{112}).
\end{eqnarray*}
 is $0.9194$ with the closest product state
 \[
\ket{\phi_\Psi} = \ket{\phi_1}\otimes\ket{\phi_2}\otimes\ket{\phi_3},\]
where
\begin{eqnarray*}
  \ket{\phi_1} &=& (0.2887 + 0.1283 i)\ket{0}+( -0.1999 - 0.9275 i)\ket{1}, \\
  \ket{\phi_2} &=&  (-0.0366 - 0.3138 i)\ket{0}+( -0.6964 + 0.6443 i)\ket{1},\\
    \ket{\phi_1} &=& (0.5420 - 0.2983 i)\ket{0}+ (-0.4013 - 0.1367 i)\ket{1} \\
               & & +(-0.5 + 0.4331 i)\ket{2}.
\end{eqnarray*}

{\it Example 9: $2\times3\times3$  system.} In this case, the theoretical upper bound is $1.0879$. It can be verified that  the GME of the symmetric pure state
\begin{eqnarray*}
\ket{\Psi} = \frac{1}{\sqrt{6}}(\ket{000}+\ket{101}+\ket{012}+\ket{110}+\ket{021}+\ket{122})
\end{eqnarray*}
is $0.9194$ with  the closest product state
\[
\ket{\phi_\Psi} =\ket{\phi_1}\otimes\ket{\phi_2}\otimes\ket{\phi_3},
\]
where
\begin{eqnarray*}
  \ket{\phi_1} &=& (0.5021 - 0.4979 i)\ket{0}+(0.1802 + 0.6838 i)\ket{1},\\
  \ket{\phi_2} &=& (0.5208 - 0.2491 i)\ket{0}+( -0.4762 - 0.3265 i)\ket{1} \\
    & & + (-0.04464 + 0.5756 i)\ket{2},\\
  \ket{\phi_3} &=& (0.1944 + 0.5437 i)\ket{0}+( 0.3736 - 0.4401 i)\ket{1} \\
   & & + (-0.5680 - 0.1035 i)\ket{2}.
\end{eqnarray*}

{\it Example 10: $2\times2\times4$  system.} In this case, the theoretical upper bound is $1$. It can be verified that  the GME of the following pure state
\begin{eqnarray*}
\ket{\Psi} = \frac{1}{2}(\ket{000}+\ket{011}+\ket{102}+\ket{113})
\end{eqnarray*}
is $1$ with  the closest product state
\[
\ket{\phi_\Psi} =\ket{\phi_1}\otimes\ket{\phi_2}\otimes\ket{\phi_3},
\]
where
\begin{eqnarray*}
  \ket{\phi_1} &=& (0.0969 - 0.7218i)\ket{0}+(0.4724 - 0.4964i)\ket{1},\\
  \ket{\phi_2} &=& (0.4498 + 0.0562i)\ket{0}+(0.8197 + 0.3501i)\ket{1}, \\
  \ket{\phi_3} &=& (0.0842 + 0.3192i)\ket{0}+(0.3321 + 0.5578i)\ket{1}\\
  && + ( 0.2404 + 0.1967i)\ket{2}+(0.5610 + 0.2416i)\ket{3}.
\end{eqnarray*}
The theoretical upper bound is reached by this particular state.


{\it Remark 3.} The results in Examples 8-11 are summarized in Table \ref{table:heterogeneous}.  It is clear that, by adding one or two qubits in a right way, the degree of entanglement might be significantly increased. More discussions on the entanglement structure of $2\times m \times n$ systems are given in  \cite{CMS10}.  A more detailed discussion of the structure of multipartite entanglement  can be found  in, e.g., \cite{HV13}. 

\begin{table}[htp]
\caption{}
\begin{center}
\begin{tabular}{|c|c|c|}
\hline
$3$-party  & theoretical upper bound & best GME found\\
\hline
$2\times 2 \times 2$  & 1                     & 0.8165\\
\hline
$2\times 2 \times 3$  & 1                 & 0.9194\\
\hline
$2\times 3 \times 3$   &   1.0879                & 0.9194\\
\hline
$2\times 2 \times 4$   &   1               & 1\\
\hline
\end{tabular}
\end{center}
\label{table:heterogeneous}
\end{table}%

{\it Example 11: $3\times3\times3\times3\times3\times2$ system.} In this case,    the theoretical upper bound is $1.3575$.  The GME of the following 2-uniform state, \cite[(31)]{GBZ16},
\begin{eqnarray*}
\ket{\Psi} &=&\frac{1}{3\sqrt{2}}(\ket{000000}+\ket{001121}+\ket{010220}\\
&& + \ket{012011}+\ket{021210}+\ket{022101}\\
&&+\ket{111110}+\ket{112201}+\ket{121000}\\
&&+\ket{120121}+\ket{102020}+\ket{100211}\\
&&+\ket{222220}+\ket{220011}+\ket{202110}\\
&&+\ket{201201}+\ket{210100}+\ket{211021})
\end{eqnarray*}
is $1.2364$    with the closest product state
\[
\ket{\phi_\Psi} =\ket{\phi_1}\otimes\ket{\phi_2}\otimes\ket{\phi_3}\otimes\ket{\phi_4}\otimes\ket{\phi_5}\otimes\ket{\phi_6},
\]
where
\begin{eqnarray*}
  \ket{\phi_1} &=& (-0.99876 - 0.0497913 i)\ket{1}, \\
  \ket{\phi_2} &=& (-0.956069 -  0.293143 i)\ket{0}, \\
  \ket{\phi_3} &=& (0.413005 - 0.910729 i)\ket{0},\\
  \ket{\phi_4} &=& (-0.739477 +  0.673182 i)\ket{2}, \\
  \ket{\phi_5} &=& (-0.99998 +  0.00639697 i)\ket{1}, \\
  \ket{\phi_6} &=& (0.028172 + 0.999603 i)\ket{1}. \\
\end{eqnarray*}
From the above, it is fair to say that the upper bound of entanglement for this class of heterogeneous systems is reasonably tight.

\section{Concluding remarks}

In this paper we have been concentrated  on the pure-state setting. This is without loss of generality, as a mixed state is a convex combination of pure states, the largest distance, in terms of geometric measure of entanglement, is always achieved by pure states.

The problem of computation of the geometric measure of entanglement is equivalent to the problem of best rank-one tensor approximation. Many algorithms in the literature of tensor computation have been proposed, see, e.g.,  \cite{NQB14, NW14, HQZ16, NB16, Teng17}, and Matlab toolboxes \cite{BG15, VDS+16}.

In the literature, several theoretical upper bounds for entanglement in terms of various measures have been derived.  For example, for any  $n$-partite system whose subsystems being $m\geq 2$ levels each, it is found \cite{CHS00} that each pure state can be associated to a tensor  with at most $m^n -nm(m-1)/2$ nonzero elements.  Based on this, an upper bound of entanglement of this class of multipartite systems in terms of Schmidt measure (the Hartley entropy)  has been proposed  in \cite{EWZ17}. However, the tightness of the upper bound is not discussed. Upper bounds for local entanglement has been discussed in \cite{LVE03}. To be more specific,  given an $n$-partite system, localizable  entanglement $E_{\rm loc}(A,B)$ between subsystems $A$ and $B$ quantifies the maximal amount of entanglement between $A$ and $B$ after performing all possible measurements on the other $n-2$ subsystems. An upper bound for  localizable  entanglement $E_{\rm loc}(A,B)$ is given in \cite{LVE03} for a 4-qubit system by means of entanglement of assistance. Finally, for a spin-1/2 chain, let $C_{ij}$ be the quantum concurrence of the subsystem composed of spins on sites $i$ and $j$, and let $\tau_{1,i}$ be the one-tangle of the spin on site $i$. Then the monogamy inequality  $\sum C_{ij}^2\leq \tau_{1,i}$ has been proved. \cite{CKW00,OV06}.  In particular, the equality holds for the 3-qubit W state \eqref{july14_W}.

\bigskip

{\bf Acknowledgment} The research is partially supported by the Hong Kong Research Grant Council,  the National Natural Science Foundation of China, and the Research Programme of National University of Defense Technology.  The authors are grateful to Shenglong Hu and Minru Bai for helpful  discussions.

\section*{Appendix}


Given a non-symmetric pure state $\ket{\Psi}\in \mathbb{C}^{d_1}\otimes\cdots\otimes \mathbb{C}^{d_n}$, denote the  corresponding tensor by $\A$, as given in \eqref{eq:Psi}. As discussed in the paper,   the maximal overlap $\langle \Psi | \phi_\Psi \rangle$ in \eqref{near3}  is equal to the largest U-eigenvalue of  the non-symmetric tensor $\A$. The following algorithm can be used to find the largest U-eigenvalue \cite{NQB14}. In fact, this algorithm computes the U-eigenvalue of  $\bar{A}$, namely, the complex conjugate of the tensor $\A$. But this is not a problem as $\A$ and $\bar{A}$ have the same  U-eigenvalues. 

{\bf Step 1 (Initial step)}: Choose a starting point $x^{(i)}_0\in \mathbb{C}^{d_i}$ with $||x^{(i)}_0||=1$, $i=1, 2, \cdots, n$. Choose a positive real number $\alpha$. Let $\lambda_0=\bar{\mathcal{A}} x^{(1)}_0 x^{(2)}_0 \cdots x^{(n)}_0$, where the operation between the tensor $\A$ and the vectors $x_0^{(i)}$ has been defined in \eqref{eq:tensor_vector}.

{\bf Step 2 (Iterating step)}: {\bf for} $k=1,2,\cdots$, do

\hskip 1.0 in {\bf for} $i=1,2,\cdots, n$, do
 
\[\hat{x}^{(i)}_{k} = \lambda_{k-1} \mathcal{A} \bar{x}^{(1)}_{k-1}\cdots \bar{x}^{(i-1)}_{k-1}\bar{x}^{(i+1)}_{k-1}\cdots \bar{x}^{(n)}_{k-1}+\alpha x^{(i)}_{k-1},
\]
and
\[
 x^{(i)}_{k}= \hat{x}^{(i)}_k/|| \hat{x}^{(i)}_k || .
 \]

\hskip 1in {\bf end for $i$}

$ \lambda_k =  \bar{\mathcal{A}} x^{(1)}_{k} x^{(2)}_{k} \cdots x^{(n)}_{k} $.

\hskip 0.8 in {\bf end for $k$}

{\bf Step 3 (Return)}:

\hskip 0.5 in U-eigenvalue $\lambda=|\lambda_k|$

\hskip 0.5 in U-eigenvector $u^{(i)}=(\frac{\lambda}{\lambda_k})^{1/d_i} x^{(i)}$, $i=1, \cdots, n$.

%
%
%






\bibliographystyle{apsrev4-1}
\bibliography{QZbib}

\end{document}